\documentclass[sigconf]{acmart}

\usepackage{tikz}
\usepackage{float}
\usetikzlibrary{positioning, arrows.meta}
\usepackage{graphicx}
\usepackage{booktabs}

\AtBeginDocument{%
  }

\setcopyright{acmlicensed}
\copyrightyear{2026}
\acmYear{2026}
\renewcommand\footnotetextcopyrightpermission[1]{}
\acmConference[ADKDD '26]{ADKDD 2026}{August 10, 2026}{Jeju, Korea}

\begin{document}

\title{Attributed, But Not Incremental: Cannibalization-Corrected Attribution for Large-Scale Advertising}

\author{Donghui Li}
\authornote{Both authors contributed equally to this research.}
\affiliation{%
  \institution{TikTok}
  \city{San Jose}
  \state{CA}
  \country{USA}
}
\email{donghui.li@bytedance.com}

\author{Bowen Yuan}
\authornotemark[1]
\affiliation{%
  \institution{TikTok}
  \city{Shanghai}
  \country{China}
}
\email{xiangyang.yuan@bytedance.com}

\author{Zili Yang}
\affiliation{%
  \institution{TikTok}
  \city{Beijing}
  \country{China}}
\email{zili@bytedance.com}

\author{Qinxin Chen}
\affiliation{%
  \institution{TikTok}
  \city{San Jose}
  \state{CA}
  \country{USA}
}
\email{qinxin.chen@bytedance.com}

\author{Lijing Song}
\affiliation{%
  \institution{TikTok}
  \city{San Jose}
  \state{CA}
  \country{USA}}
\email{leonasu@bytedance.com}

\renewcommand{\shortauthors}{Li et al.}

\begin{abstract}
In large-scale paid acquisition and growth advertising systems, production attribution outputs are widely used for daily budget allocation and channel diagnosis. However, paid-attributed conversions such as daily new users (DNU) may systematically overstate true incremental growth when paid channels overlap with organic demand, brand-driven traffic, or other acquisition channels. This attribution-cannibalization mismatch can distort incremental ROI measurement and budget decisions at scale.

We propose an experiment-calibrated attribution correction framework that uses incrementality experiments as causal anchors to convert sparse lift measurements into daily correction estimates. To make the corrected signal actionable at production granularity, we further allocate calibrated cannibalization volume across business hierarchies under structural consistency constraints. Offline forward-in-time validation against channel-level incrementality experiment readouts shows that the proposed framework substantially reduces calibration error relative to raw attribution and fine-grained ML baselines. Deployed across multiple global TikTok markets, the system supported budget and traffic strategy adjustments that were followed by an approximately 15-percentage-point reduction in the measured cannibalization rate.

\end{abstract}



\maketitle

\section{Introduction}

Paid acquisition systems in content platforms, e-commerce, and other growth businesses allocate substantial budgets across multiple advertising channels. These decisions are typically guided by attribution, which serves as the measurement layer for conversions such as daily new users (DNU) and informs ROI estimation, channel diagnosis, budget reallocation, and campaign evaluation. In production, this layer often relies on last-touch attribution (LTA) or simplified data-driven attribution (DDA), which assign credit based on observed user paths. The credibility of downstream decisions therefore depends on whether the channel receiving credit is also close to the channel creating incremental value.

\begin{figure}[t]
\centering
\includegraphics[width=\columnwidth]{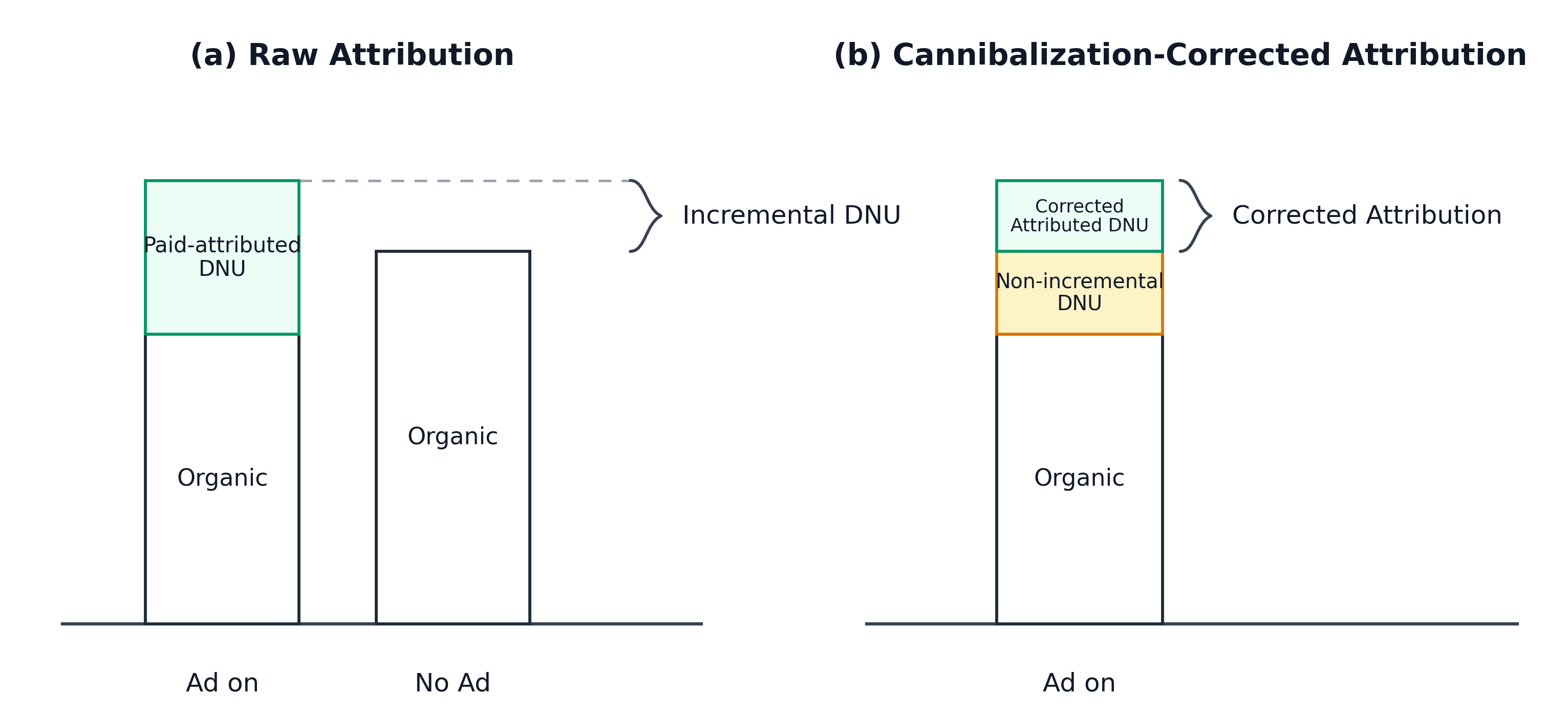}
\caption{Attribution-cannibalization mismatch and correction. Panel (a) shows the non-incremental excess \(H_t = A_t - Lift_t\); panel (b) shows experiment-calibrated correction toward incrementality-aligned attribution. Schematic, not drawn to scale.}
\Description{
Schematic bar chart showing that raw paid attribution can exceed experimental lift, and that corrected attribution removes the non-incremental component.
}
\label{fig:cannibalization_concept}
\end{figure}

In practice, this assumption often fails. When paid channels
overlap with organic demand, brand-driven traffic, or other active acquisition channels, attribution can
assign credit to users who would have converted even without
the focal paid intervention. We refer to this gap between credited conversions and causal incremental conversions as the \emph{attribution--cannibalization mismatch}, as illustrated in Figure~\ref{fig:cannibalization_concept}(a). In our evaluation across multiple global markets in a large-scale content platform, paid-attributed DNU consistently exceeds experimentally measured incremental DNU, indicating that a material share of nominal paid attribution is non-incremental. This mismatch rewards channels that harvest existing demand rather than those that create new demand.

This mismatch matters because attribution sits upstream of nearly every paid-media decision. Errors at the measurement layer propagate into ROI estimates, budget allocation, and campaign evaluation, making low-incrementality traffic appear efficient and diverting spend away from channels with stronger causal impact. The problem is especially challenging in production because the available signals have complementary weaknesses. Production attribution is timely, granular, and continuously available, but observational by construction. Incrementality experiments provide credible counterfactual lift estimates, but they are sparse, delayed, costly, and infeasible to run continuously across all channels, markets, and campaign hierarchies. The practical measurement problem is therefore not whether to replace attribution with experiments, but how to use experimental evidence to calibrate attribution into an incrementality-aligned operating metric, as illustrated in Figure~\ref{fig:cannibalization_concept}(b).

We therefore propose an experiment-calibrated cannibalization correction framework that combines the causal credibility of incrementality experiments with the coverage of production attribution: \textbf{Experiment-to-Daily Cannibalization + Hierarchical Cannibalization Allocation (ETDC+HCA)}. The framework first uses experimental lift estimates to anchor the scale of channel-level cannibalization, and then extrapolates these corrections to daily channel-level attribution through proxy-variable modeling. A hierarchical allocation layer then distributes calibrated cannibalization volume to finer business units while enforcing hierarchy-consistent aggregation. The resulting corrected attribution preserves the timeliness and granularity required for production decision-making while aligning reported conversions more closely with experimentally measured incremental lift.

This paper makes three contributions. First, we formulate cannibalization correction as a causal calibration problem for production attribution in paid acquisition and growth advertising systems, clarifying the distinction between credited conversions and incremental conversions. Second, we propose an experiment-to-daily cannibalization layer that converts sparse channel-level lift signals into daily incrementality-aligned correction estimates. Third, we introduce a constrained hierarchical allocation layer and validate the full framework through forward-in-time evaluation against channel-level incrementality experiments and production deployment across multiple global markets.

\section{Overview of Related Work}

\noindent\textit{Attribution.}
Prior work on attribution has evolved from heuristic rules to data-driven and sequential models. Early multi-touch attribution methods estimate channel contributions from observed user paths rather than assigning all credit to first- or last-touch events \cite{shao2011}. Later systems incorporate temporal structure and richer model classes, including RNN- and Shapley-based causal MTA \cite{du2019}, deep attribution models \cite{li2018dnamta,yang2020}, and large-scale industrial data-driven attribution (DDA) systems \cite{bencina2025}. These methods improve journey-level credit allocation but remain observational. We instead treat production attribution as the operational signal and calibrate it with experimental lift, then propagate the correction through a business hierarchy.

\noindent\textit{Incrementality.}
Incrementality experiments provide a more credible basis for causal measurement. Large-scale Facebook field experiments show that observational ad measurement can diverge substantially from randomized experiment estimates \cite{gordon2019}. Prior work argues that attribution should be evaluated against counterfactual outcomes identified through randomized experiments rather than prediction accuracy alone \cite{dalessandro2012}. However, advertising experiments are costly, delayed, and statistically noisy, especially when effects are small or heterogeneous \cite{lewis2015}. Designs such as Ghost Ads and controlled geo-experiments improve the feasibility of online ad measurement, but still require controlled traffic or market-level interventions and cannot provide continuous fine-grained coverage for production decision-making \cite{johnson2017,barajas2020}. Experiments are therefore essential causal anchors, but they are too sparse to replace production attribution.

\noindent\textit{Cannibalization.}
Cannibalization is one important reason attributed conversions can diverge from incremental conversions. Large-scale paid search experiments show that ads may capture users who would otherwise arrive organically, causing ad-attributed outcomes to overstate true incremental value \cite{blake2015}. At the same time, channel interactions are not always purely substitutive: paid channels may also generate complementary organic activity or spillovers \cite{ju2025}. This makes cannibalization a counterfactual problem rather than a correlation problem: the key quantity is what would have happened without the focal paid intervention \cite{aguilar2021}.

\noindent\textit{Research gap.}
Our work connects these three lines of research in a production attribution system. Attribution provides timely and granular signals but lacks causal scale; experiments provide causal lift but are sparse and delayed; cannibalization explains a systematic gap between the two. We therefore treat production attribution as an operational measurement layer and use incrementality experiments to calibrate it toward incremental DNU. Unlike prior work focused mainly on modeling observed journeys or measuring isolated lift, we provide a deployable mechanism that transforms sparse experimental signals into daily, hierarchy-consistent attribution corrections that support budget allocation and channel diagnosis at scale. 

\vspace{1pt}
\section{Proposed Approach}
\subsection{Problem Formulation}
In large-scale advertising and user growth systems, attribution-related measurement relies on two complementary information sources. Production attribution pipelines continuously report paid-attributed conversions at fine granularity with negligible delay, but they are observational and do not identify counterfactual channel contribution. Incrementality experiments identify counterfactual lift more directly, but their readouts are delayed and cover only a sparse subset of channels and time periods. The instant signal is causally biased; the causal signal is delayed and sparse. We formalize how to combine the two so that production attribution can be calibrated toward experimentally identified incremental lift.

We focus on daily new users (DNU) as the primary conversion outcome. Let \(A_t\) denote nominal paid-attributed DNU at time \(t\), available from the production attribution system. Let \(Y_t^{ad}\) denote total conversions under the realized advertising exposure and \(Y_t^{no\text{-}ad}\) denote the counterfactual total conversions without that exposure. The true incremental lift is:
\begin{equation}
    Lift_t = Y_t^{ad} - Y_t^{no\text{-}ad}.
\end{equation}

For periods with \(A_t > 0\), we define the cannibalization rate as the fraction of nominal paid-attributed conversions that is not truly incremental:
\begin{equation}
C_t = 1 - \frac{Lift_t}{A_t}.
\end{equation}

The corresponding implied cannibalization gap is:
\begin{equation}
H_t = A_t - Lift_t = A_t C_t.
\end{equation}
Positive values of \(H_t\) represent the non-incremental portion of nominal paid attribution.

Given an estimated cannibalization rate \(\hat{C}_t\), the corrected incremental attribution volume is:
\begin{equation}
    \hat{I}_t = A_t(1-\hat{C}_t).
\end{equation}



Because \(C_t\) is not directly observable in production, it must be estimated from limited experimental signals. Negative estimates may arise when measured lift exceeds nominal attribution, indicating potential organic spillover, channel complementarity, or missing attribution touchpoints. In production, we constrain \(\hat{C}_t\) within a pre-specified operational range and treat negative estimates as diagnostic signals.

This formulation leads to two challenges: attribution bias, where \(A_t\) can systematically overstate \(Lift_t\), and an experimental generalization gap, where lift is only observed at coarse granularity and in limited time windows. Our goal is to estimate a stable and scalable approximation \(\hat{C}_t\), and generalize it to fine-grained decision units to calibrate nominal paid attribution toward true incremental conversions.

\subsection{Framework Overview}

\begin{figure}[t]
\centering
\includegraphics[width=\columnwidth]{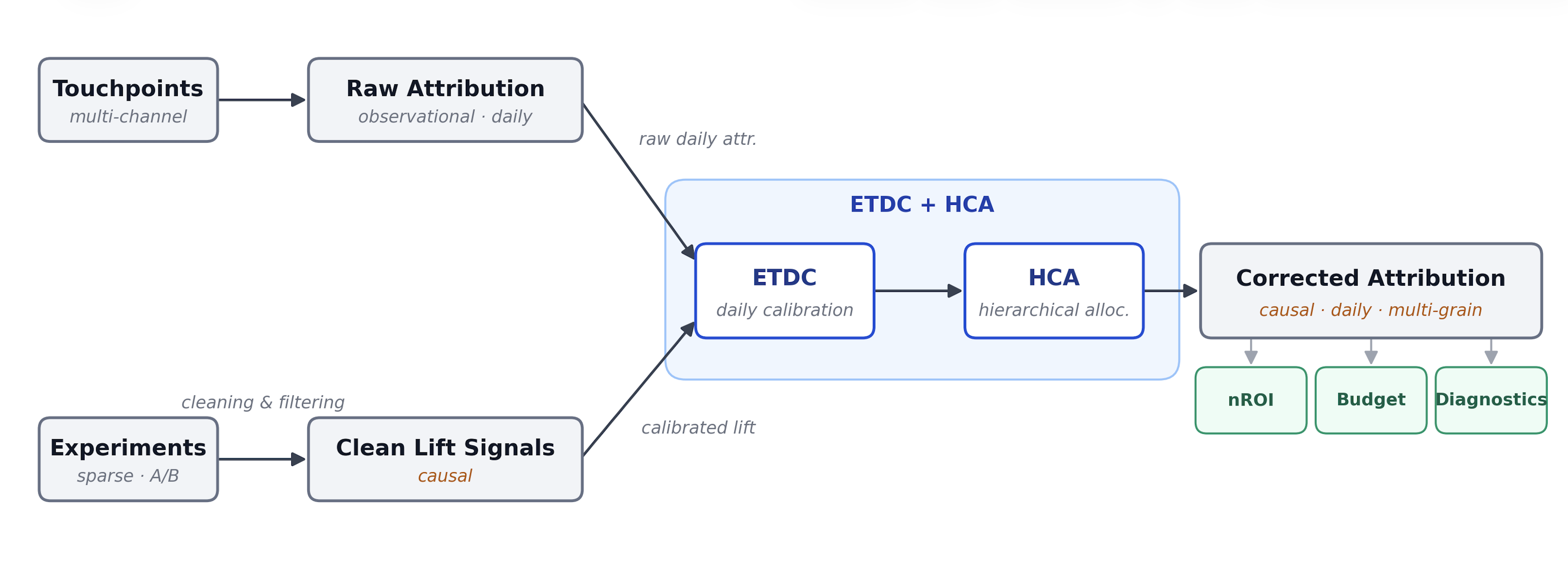}
\caption{Overview of the proposed ETDC+HCA framework.}
\Description{Overview diagram showing production attribution, experiment calibration, cannibalization correction, corrected attribution, and downstream decision outputs.}
\label{fig:framework_overview}
\end{figure}

Building on the problem formulation in Section~3.1, Figure~\ref{fig:framework_overview} summarizes the proposed \textbf{ETDC+HCA} framework, where ETDC denotes \textbf{Experiment-to-Daily Cannibalization} and HCA denotes \textbf{Hierarchical Cannibalization Allocation}. The framework combines two complementary inputs. Production attribution provides timely, fine-grained attributed DNU with broad channel and campaign coverage, but it is observational and not necessarily aligned with true incrementality. Incrementality experiments provide sparse and delayed but causally grounded lift signals, which serve as the calibration anchor for correction.

After signal cleaning, pull-forward adjustment for accelerated conversions, and quality filtering, raw experiment outputs are converted into clean lift signals. The correction layer contains two cascaded components. First, ETDC extends sparse experimental signals to continuous daily channel-level cannibalization estimates using proxy-variable modeling. Second, HCA allocates the calibrated cannibalization volume through a predefined business hierarchy, preserving aggregate consistency with experimental lift estimates while producing fine-grained correction outputs.

The framework outputs incrementality-aligned attribution at daily and multi-granularity production views, supporting downstream net incremental ROI (nROI) evaluation, budget allocation, and attribution diagnostics. Importantly, ETDC+HCA operates as a lightweight add-on layer without modifying the underlying production attribution pipeline, reducing deployment cost and operational risk.

\vspace{1pt}
\subsection{Experiment Signal Construction}

Incrementality experiments serve as the causal calibration anchors of our framework. We standardize heterogeneous experiment outputs into cleaned channel-day lift signals by aligning channel definitions, experiment windows, and conversion metrics under a unified measurement scope. We then apply standardized quality controls to exclude observations affected by insufficient delivery, abnormal blocking, configuration changes, or cross-experiment interference. Pull-forward effects are corrected using post-stabilization windows when available and otherwise adjusted with pre-defined coefficients. Experiment durations are set to ensure adequate qualified sample sizes for stable model training and calibration; experiments with insufficient sample coverage are excluded from the calibration set.

After preprocessing, we obtain a qualified channel-day sample set \(\mathcal{E}\). For each \((c,t)\in\mathcal{E}\), \(Lift_{c,t}\) denotes the cleaned experiment-derived lift signal after scope alignment, pull-forward treatment, and quality filtering. These signals provide the noisy but causally anchored supervision for experiment-to-daily cannibalization in Section~3.4.

\subsection{Experiment-to-Daily Cannibalization}

Under the formulation in Section~3.1, true cannibalization rates cannot be continuously observed in production. Channel-level experiments provide credible incremental lift, but they are sparse and delayed. Production attribution, in contrast, is dense and timely but miscalibrated. The central challenge is therefore not merely to predict daily lift, but to transfer causal scale from sparse, delayed experiments to continuously available production signals. We formulate ETDC as an experiment-to-daily cannibalization problem: learning a stable mapping from continuously observed channel-day features to calibrated lift, so that delayed experimental evidence can be converted into timely daily correction estimates.

For each channel-day pair \((c,t)\), the production attribution system provides nominal paid-attributed DNU \(A_{c,t}\). For the subset of channel-day pairs covered by qualified experiments, we additionally observe channel-level daily lift \(\mathrm{Lift}_{c,t}\). Because daily experimental lift can be unstable due to sample size, conversion rate, and short-term traffic fluctuations, ETDC treats it as noisy causal supervision rather than a noise-free daily label.

The ETDC feature design follows a single principle: continuously available inputs should explain movements in the organic baseline. Accordingly, for each channel-day pair \((c,t)\), we construct a feature vector \(\phi_{c,t}=(\phi^P_{c,t},\phi^T_{c,t},\phi^S_{c,t})\) from three feature blocks. The proxy block \(\phi^P\) tracks movements in the counterfactual organic baseline; the temporal block \(\phi^T\) captures recurring calendar structure such as day-of-week effects, holidays, and seasonality; and the channel/media state block \(\phi^S\) describes delivery intensity, market conditions, and campaign lifecycle. To keep calibration interpretable, proxy variables are selected according to two criteria: relevance, meaning stable association with underlying organic growth, and approximate exogeneity, meaning that they do not mechanically respond to short-term campaign switches, budget changes, or attribution rule updates.

The model learns the following mapping:
\begin{equation}
    \widehat{Lift}_{c,t}=f(\phi_{c,t};\theta).
\end{equation}

Rather than fixing a specific predictor class for \(f(\cdot)\), we design this layer to remain scale-consistent with experimental lift while satisfying production requirements for stability, interpretability, and maintainability. In deployment, we use a generalized linear model for robustness and operational simplicity. The contribution lies not in the predictor architecture itself, but in using sparse experimental lift as calibration supervision for continuous correction of production attribution.

Using the qualified samples defined in Section~3.3, we train the model on \(\mathcal{E}\) by minimizing a robust loss:

\begin{equation}
    \min_{\theta}
\sum_{(c,t)\in \mathcal{E}}
\ell\left(
f(\phi_{c,t};\theta), Lift_{c,t}
\right).
\end{equation}

We implement \(\ell(\cdot,\cdot)\) as Huber loss to reduce the influence of extreme daily experimental points that remain after input filtering. In this design, experimental lift provides both causal scale and temporal trend information, while input filtering and robust loss mitigate high variance and residual noise in daily experimental supervision.

Given \(\widehat{Lift}_{c,t}\), we recover the channel-day cannibalization rate and corrected incremental attribution using the definitions in Section~3.1:
\begin{equation}
    \widehat{C}_{c,t}
=
1-\frac{\widehat{Lift}_{c,t}}{A_{c,t}},
\qquad
\widehat{I}_{c,t}
=
A_{c,t}(1-\widehat{C}_{c,t}).
\end{equation}

In this way, incrementality that is directly observable only for experiment-covered samples is extended to a production-wide channel-day correction surface within the supported attribution scope. This layer does not further allocate corrections to device, placement, campaign, creative, or other finer-grained units; the downstream module performs fine-grained allocation while preserving channel-level aggregate consistency.

In production, we apply temporal smoothing and boundary constraints to the predicted outputs. These constraints are not intended to change the causal scale identified by experiments, but to suppress unnecessary volatility caused by daily experimental noise or short-term abnormal fluctuations, thereby improving the operational usability of the correction surface. Negative cannibalization estimates are treated as diagnostic signals for potential organic spillover, channel complementarity, or missing attribution touchpoints, and can be bounded in reporting or budget systems according to business requirements.

\subsection{Hierarchical Cannibalization Allocation}

Section~3.4 produces ETDC-calibrated cannibalized volume at a coarse granularity, such as channel-day. However, operational decisions are often made at finer business units, including campaigns, placements, devices, and their combinations. HCA does not re-estimate causal lift at these finer levels. Instead, it propagates the ETDC-calibrated total through a predefined business hierarchy, producing actionable fine-grained corrections while satisfying three operational constraints: aggregate consistency, feasibility, and locality. Specifically, child allocations sum to the calibrated parent total, allocated cannibalization remains non-negative and does not exceed nominal attributed volume, and local changes mainly affect the corresponding subtree rather than unrelated branches.

Let \(\widehat{H}_{r,t}^{(0)}\) denote the ETDC-calibrated root cannibalized volume at time \(t\). For any parent node \(g\) at layer \(l-1\), let \(\widehat{H}_{g,t}^{(l-1)}\) denote its calibrated cannibalized volume, and let \(\mathcal{C}(g)\) denote its child set. HCA distributes the fixed parent total to its children under two constraints:
\begin{equation}
    \sum_{j\in\mathcal{C}(g)}
\widehat{H}_{j,t}^{(l)}
=
\widehat{H}_{g,t}^{(l-1)},
\qquad
0\leq
\widehat{H}_{j,t}^{(l)}
\leq
A_{j,t}^{(l)}.
\end{equation}

In production, HCA is applied only to positive calibrated cannibalization volume. Negative cannibalization values are treated separately as diagnostic signals and are not propagated through the main hierarchical allocation path.

To distribute a fixed parent total within a sibling set, HCA computes a relative cannibalization propensity for each child node:
\begin{equation}
    s_{j,t}^{(l)}=q^{(l)}(z_{j,t}^{(l)};\psi^{(l)}),
\end{equation}
where \(z_{j,t}^{(l)}\) denotes stable local features at layer \(l\). These scores are not interpreted as child-level causal effects or absolute cannibalization rates; they serve only as allocation priors for distributing the ETDC-calibrated parent total.

Sibling propensities are normalized into allocation weights, which determine child-level cannibalized volume:
\begin{equation}
\widetilde{w}_{j,t}^{(l)}
=
\frac{s_{j,t}^{(l)}}
{\sum_{k\in\mathcal{C}(g)}s_{k,t}^{(l)}},
\qquad
\widehat{H}_{j,t}^{(l)}
=
\widetilde{w}_{j,t}^{(l)}
\widehat{H}_{g,t}^{(l-1)}.
\end{equation}
If a child reaches its feasibility bound, residual volume is redistributed within the same sibling set. The corrected incremental attribution at each node is
\begin{equation}
\widehat{I}_{j,t}^{(l)}
=
A_{j,t}^{(l)}
-
\widehat{H}_{j,t}^{(l)}.
\end{equation}

Because redistribution is performed only within each parent's child set, HCA preserves locality: local policy or traffic changes primarily perturb the corresponding subtree rather than unrelated branches. The resulting fine-grained outputs should therefore be interpreted as operational allocations under ETDC-calibrated aggregate constraints, rather than independently identified unit-level causal lift estimates.

\section{Offline Experiments}
\subsection{Evaluation Design}

We follow a forward-in-time rolling protocol. For each model version, incrementality experiments available before time \(T\) are used to construct calibration signals, while subsequent experiment readouts are used for evaluation at the same channel and experiment-window granularity. After evaluation, these readouts can enter the next calibration cycle. Thus, the same experiment is not used both for calibration and evaluation within the same model version.

For each experiment \(e\), let \(Lift^{exp}_e\) denote the experiment-estimated incremental DNU, and let \(\widehat{Lift}^{(m)}_e\) denote the prediction from method \(m\), aggregated to the same channel and experiment window. We report absolute relative error (ARE) for aggregate calibration:
\begin{equation}
\mathrm{ARE}^{(m)}_e
=
\left|
\frac{
\widehat{Lift}^{(m)}_e - Lift^{exp}_e
}{
Lift^{exp}_e
}
\right|.
\end{equation}
For slice-level diagnostics, we also report signed relative error (RE), which indicates whether a method overestimates or underestimates experimental lift:
\begin{equation}
\mathrm{RE}^{(m)}_e
=
\frac{
\widehat{Lift}^{(m)}_e - Lift^{exp}_e
}{
Lift^{exp}_e
}.
\end{equation}
When aggregating across experiments, we use lift-weighted averages to reduce instability from low-lift experiments.

We compare three methods at the channel-experiment granularity. \textbf{Raw Attribution} uses production attribution without correction. \textbf{Device ML} estimates organic-like propensity at the device or record level without experiment-to-daily cannibalization. \textbf{ETDC+HCA} is the full proposed framework, combining experiment-to-daily cannibalization with hierarchical cannibalization allocation. Since HCA preserves the calibrated channel-level total by construction, channel-level ARE mainly evaluates the calibration layer; we assess HCA's operational value through the locality case study in Section~4.4.

\subsection{Overall Performance}

This section summarizes the overall performance across 18 rounds of channel-level A/B incrementality experiments, covering up to eight markets.

In the underlying experiment comparison, Raw Attribution substantially overestimates incremental contribution, confirming that uncorrected attribution contains a large non-incremental component. To protect commercially sensitive information, Table~\ref{tab:overall_performance} reports normalized error indices and relative error reduction rather than absolute lift ratios or raw error magnitudes.

Device ML reduces normalized calibration error by 69.11\% relative to Raw Attribution, but its signed-error distribution remains relatively wide across experiment slices. ETDC+HCA achieves the best overall calibration, reducing normalized calibration error by 91.38\%, with median signed error close to zero and a narrower interquartile range. These results show that the proposed framework substantially narrows the gap between production attribution and experimentally measured incremental lift while maintaining stable calibration across slices.

\begin{table}[t]
\centering
\caption{Overall performance against channel-level incrementality experiments.}
\label{tab:overall_performance}
\scriptsize
\setlength{\tabcolsep}{2.5pt}
\resizebox{\columnwidth}{!}{%
\begin{tabular}{@{}lcccc@{}}
\toprule
Method & Error Index & Err. Red. & Med. RE & IQR \\
\midrule
Raw Attr. & 1.00 & --      & --       & -- \\
Device ML & 0.31 & 69.11\% & -4.80\%  & [-38.08\%, 44.27\%] \\
ETDC+HCA  & 0.09 & 91.38\% & 0.60\%   & [-8.11\%, 7.24\%] \\
\bottomrule
\end{tabular}%
}
\par\vspace{0.8em}
\begin{minipage}{\columnwidth}
\footnotesize
Error Index is normalized by the absolute relative error of Raw Attribution; lower is better.
Err. Red. is relative to Raw Attribution; Med. RE is median signed relative error.
IQR reports the 25th--75th percentile range of signed relative error.
Raw Attribution signed-error distribution and absolute magnitudes are omitted for confidentiality.
ETDC-only and ETDC+HCA have identical channel-level metrics because HCA preserves calibrated channel totals; HCA is evaluated in Section~4.4.
\end{minipage}
\vspace{3pt}
\end{table}

\subsection{Performance Breakdown}

To further illustrate model behavior at finer evaluation slices, Table~\ref{tab:performance_breakdown} reports representative results from two markets across two channels and two experiment rounds.

Raw Attribution consistently overestimates incremental contribution in these slices, with signed relative errors ranging from 179\% to 334\%. Device ML reduces error in some cases but remains unstable, alternating between overestimation and substantial underestimation across market-channel-round slices. In contrast, ETDC+HCA stays close to experimental lift across all reported slices, with signed relative errors ranging from -7\% to 10\%. This indicates that the overall gain is not driven by aggregate averaging alone; ETDC+HCA also maintains more stable calibration across heterogeneous evaluation slices.

\begin{table}[t]
\centering
\caption{Performance breakdown on representative market-channel experiment slices. Values are signed relative errors against experimental lift estimates.}
\small
\label{tab:performance_breakdown}
\begin{tabular}{@{}lllrrr@{}}
\toprule
Market & Channel & Round & ETDC+HCA & Device ML & Raw Attr. \\
\midrule
A & Channel 1 & Round 1 & -2\% & 58\% & 214\% \\
A & Channel 1 & Round 2 & -7\% & 46\% & 179\% \\
A & Channel 2 & Round 1 & 3\% & -63\% & 200\% \\
A & Channel 2 & Round 2 & -4\% & -64\% & 225\% \\
B & Channel 1 & Round 1 & 10\% & 11\% & 214\% \\
B & Channel 1 & Round 2 & -4\% & 33\% & 224\% \\
B & Channel 2 & Round 1 & 7\% & -63\% & 334\% \\
B & Channel 2 & Round 2 & 3\% & -75\% & 259\% \\
\bottomrule
\end{tabular}

\vspace{0.3em}
\footnotesize{
Signed relative error is computed as
(predicted lift - experimental lift estimate) / experimental lift estimate.
Positive values indicate overestimation and negative values indicate underestimation. Full per-slice results follow the same pattern and are omitted for brevity.
}
\vspace{3pt}
\end{table}

\subsection{Operational Locality Validation}
Channel-level ARE mainly evaluates aggregate calibration and does not fully capture the value of HCA, as HCA preserves the calibrated channel-level total by construction. Instead, HCA is designed to optimize how calibrated correction volumes are allocated to actionable business dimensions. We validate the property using a localized production policy update within one advertising channel. We compare the full ETDC+HCA framework against an ETDC-only baseline that applies calibrated correction without hierarchical allocation.

\begin{figure}[t]
\centering
\includegraphics[width=\columnwidth]{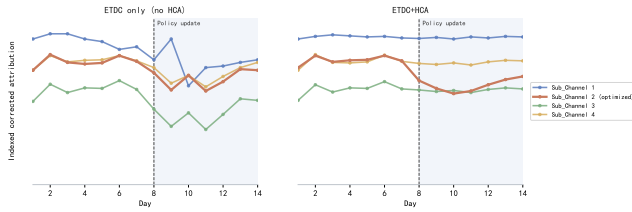}
\caption{Operational locality validation under a local policy update. Curves show indexed corrected attribution after cannibalization correction, with absolute scale anonymized. ETDC-only correction diffuses the update across multiple channels, while ETDC+HCA concentrates the change on the optimized channel and keeps unrelated channels largely stable.}
\Description{Line plots comparing ETDC-only and ETDC+HCA correction behavior before and after a localized policy update.}
\label{fig:operational_locality}
\end{figure}

As shown in Figure~\ref{fig:operational_locality}, ETDC-only correction diffuses the policy impact across multiple channels, causing unrelated channels to exhibit spurious changes after the update. In contrast, ETDC+HCA concentrates most of the attribution change on the target channel, while unrelated channels remain largely stable. This result indicates that HCA improves dimensional consistency and reduces correction spillover, making corrected attribution outputs more actionable for channel-level diagnosis and budget adjustment.

\section{Online Deployment and Impact}
The proposed method is deployed as a causal correction layer on top of the existing production attribution pipeline, rather than replacing it. The original attribution system continues to provide nominal paid-attributed DNU, while the correction layer combines experimental signals, daily features, and raw attribution outputs to produce calibrated incremental attribution. The corrected incremental attribution is written back to downstream decision modules, including nROI, budget allocation, channel diagnosis, and campaign evaluation. This design preserves the coverage and timeliness of production attribution while aligning its outputs more closely with true incrementality.

After deployment, the corrected attribution outputs were used to guide budget optimization and traffic strategy adjustments. Informed by these outputs, downstream teams adjusted budget allocation and traffic strategy; the measured overall cannibalization rate subsequently decreased by approximately 15 percentage points, consistent with the system helping surface low-incrementality traffic and redirect spend toward higher-incrementality channels.

\section{Limitations}

The framework depends on the quality and coverage of incrementality experiments. Sparse experiments, wide confidence intervals, incomplete treatment isolation, or unresolved pull-forward effects can propagate uncertainty into calibration. Proxy-based extrapolation also requires monitored relevance and approximate exogeneity; product launches, seasonality, market shocks, or acquisition-channel shifts may weaken these assumptions and require recalibration. Finally, fine-grained outputs should be interpreted as calibrated allocations rather than independently identified unit-level causal effects. Negative cannibalization, organic spillover, and channel complementarity are treated as diagnostic signals rather than fully modeled effects.

\section{Conclusion and Future Work}

We study attribution-cannibalization mismatch in large-scale paid acquisition and growth advertising systems, where paid-attributed DNU can systematically overstate true incremental growth. We propose an experiment-calibrated correction layer that uses incrementality experiments as causal anchors, extrapolates sparse lift signals into daily correction estimates, and allocates calibrated non-incremental volume through hierarchy-consistent constraints. Experiments and production deployment show that the corrected outputs better align with measured lift and provide more reliable signals for nROI measurement, channel diagnosis, and budget optimization. Future work will extend the framework toward richer cross-channel substitution, complementarity, organic spillover modeling, and automated recalibration as new experiments and market conditions evolve.

\bibliographystyle{ACM-Reference-Format}
\bibliography{reference}

\end{document}